%% file: Manuscript.tex
\documentclass[conference]{IEEEtran}

\input{tex/packages.tex}

\newif\ifdoubleblind 

\renewcommand{\baselinestretch}{0.95}

\begin{document}

\input{tex/variables.tex}
\input{tex/acro.tex}

\acresetall
\input{tex/title}
\input{tex/arxiv}
\input{tex/abstract.tex}

\IEEEpeerreviewmaketitle

\input{tex/introduction.tex}
\input{tex/relatedWork.tex}

\input{tex/solution_approach.tex}
\input{tex/methodology.tex}

\input{tex/results.tex}
\input{tex/conclusion.tex}
\input{tex/acknowledgment.tex}

\bibliographystyle{IEEEtran}
\bibliography{Bibliography}

\end{document}

%% file: tex/packages.tex
\usepackage{graphicx}
\usepackage{epstopdf}
\usepackage{standalone}
\usepackage[nolist]{acronym}
\usepackage{tcolorbox}

\usepackage{pdfpages}
\usepackage{tikz}
\usetikzlibrary{positioning}
\usepackage[bookmarks=false]{hyperref}
\usepackage{pdfcomment}
\usepackage{hologo}

\usepackage{listings}
\definecolor{codegreen}{rgb}{0,0.6,0}
\definecolor{codegray}{rgb}{0.5,0.5,0.5}
\definecolor{codepurple}{rgb}{0.58,0,0.82}
\definecolor{backcolour}{rgb}{0.9020,0.9294,0.9608}

\lstdefinestyle{mystyle}{
	backgroundcolor=\color{backcolour},   
	commentstyle=\color{codegreen},
	keywordstyle=\color{magenta},
	numberstyle=\tiny\color{codegray},
	stringstyle=\color{codepurple},
	basicstyle=\ttfamily\footnotesize,
	breakatwhitespace=false,         
	breaklines=true,                 
	captionpos=b,                    
	keepspaces=true,                 
	numbers=left,                    
	numbersep=5pt,                  
	showspaces=false,                
	showstringspaces=false,
	showtabs=false,                  
	tabsize=2
}
\usepackage{xstring}

\usepackage[utf8]{inputenc}
\usepackage{marvosym}

\usepackage{algorithm}
\usepackage{algpseudocode}
\usepackage{tikz}

\definecolor{flexicolor}{RGB}{46,49,146}
\definecolor{amaricolor}{RGB}{237,28,36}

\usepackage{xspace}

\usepackage[binary-units=true]{siunitx}
\usepackage{psfrag}
\usepackage{graphicx}
\usepackage{tabularx,booktabs}
\usepackage{multirow, makecell}
\usepackage{rotating}

\renewcommand{\baselinestretch}{1}

\usepackage{multirow}
\usepackage[cmex10]{amsmath}
\usepackage{wasysym}
\usepackage[caption=false,font=footnotesize]{subfig}
\usepackage{stfloats}
\hypersetup{bookmarks=false}

%% file: tex/variables.tex
\newcommand{\paperTitle}{LIMITS: Lightweight Machine Learning for IoT Systems with Resource Limitations}
\newcommand{\paperAuthors}{Benjamin Sliwa$^1$, Nico Piatkowski$^2$, and Christian Wietfeld$^1$}
\newcommand{\paperEmails}{$\{$Benjamin.Sliwa, Nico.Piatkowski, Christian.Wietfeld$\}$@tu-dortmund.de}
\newcommand{\framework}{\ac{LIMITS}\xspace}

\newcommand{\figurePadding}{0pt}
\newcommand{\figureTopPadding}{\figurePadding}
\newcommand{\figureBottomPadding}{\figurePadding}

\newcommand{\msp}{\texttt{MSP430}\xspace}
\newcommand{\atmega}{\texttt{ATmega328}\xspace}
\newcommand{\esp}{\texttt{ESP32}\xspace}

\newcommand{\sfw}{0.24}

\newcommand{\dummy}[3]
{
	\begin{figure}[b!]  
		\begin{tikzpicture}
		\node[draw,minimum height=6cm,minimum width=\columnwidth]{\LARGE #1};
		\end{tikzpicture}
		\caption{#2}
		\label{#3}
	\end{figure}
}

\newcommand{\wDummy}[3]
{
	\begin{figure*}[b!]  
		\begin{tikzpicture}
		\node[draw,minimum height=6cm,minimum width=\textwidth]{\LARGE #1};
		\end{tikzpicture}
		\caption{#2}
		\label{#3}
	\end{figure*}
}

\newcommand{\basicFig}[7]
{
	\begin{figure}[#1]  	
		\vspace{#6}
		\centering		  
		\includegraphics[width=#7\columnwidth]{#2}
		\caption{#3}
		\label{#4}
		\vspace{#5}	
	\end{figure}
}
\newcommand{\fig}[4]{\basicFig{#1}{#2}{#3}{#4}{0cm}{0cm}{1}}

\newcommand{\subfig}[3]%
{%
	\subfloat[#3]%
	{%
		\includegraphics[width=#2\textwidth]{#1}%
	}%
	\hfill%
}

\newcommand{\subfigh}[3]%
{%
	\subfloat[#3]%
	{%
		\includegraphics[height=3.65cm]{#1}%
	}%
	\hfill%
}

\newcommand\circled[1] 
{
	\tikz[baseline=(char.base)]
	{
		\node[shape=circle,draw,inner sep=1pt] (char) {#1};
	}\xspace
}

%% file: tex/acro.tex
\begin{acronym}
	\acro{RF}{Random Forest}
	\acro{DL}{Deep Learning}
	\acro{SVM}{Support Vector Machine}
	\acro{M5}{M5 Regression Tree}
	\acro{RBF}{Radial Basis Function}
	\acro{ANN}{Artificial Neural Network}
	\acro{CNN}{Convolutional Neural Network}
	\acro{KNN}{k-Nearest Neighbor}
	\acro{CSR}{Classification Success Ratio}
	\acro{CART}{Classification and Regression Tree}
	\acro{WEKA}{Waikato Environment for Knowledge Analysis}
	\acro{KNIME}{Konstanz Information Miner}
	\acro{MDI}{Mean Decrease Impurity}
	\acro{MAE}{Mean Absolute Error}
	\acro{RMSE}{Root Mean Square Error}
	\acro{UI}{User Interface}
	\acro{CLI}{Command Line Interface}
	\acro{CPS}{Cyber-physical System}
	\acro{WCET}{Worst-case Execution Time}
	\acro{IoT}{Internet of Things}
	\acro{ESP-IDF}{Espressif IoT Development Framework}
	\acro{RAM}{Random Access Memory}
	\acro{DDNS}{Data-driven Network Simulation}
	\acro{ITS}{Intelligent Transportation System}
	\acro{UE}{User Equipment}
	\acro{TCP}{Transmission Control Protocol}
	
	\acro{LTE}{Long Term Evolution}
	\acro{MNO}{Mobile Network Operator}
	\acro{eNB}{evolved Node B}
	
	\acro{SS}{Signal Strength}
	\acro{ASU}{Arbitrary Strength Unit}
	\acro{RSSI}{Reference Signal Strength Indicator}
	\acro{RSRP}{Reference Signal Received Power}
	\acro{RSRQ}{Reference Signal Received Quality}
	\acro{SINR}{Signal-to-interference-plus-noise Ratio}
	\acro{CQI}{Channel Quality Indicator}
	\acro{TA}{Timing Advance}
	\acro{MCU}{Microcontroller Unit}
	\acro{LIMITS}{LIghtweight Machine learning for IoT Systems}
\end{acronym}

%% file: tex/title.tex
\title{\paperTitle}

\ifdoubleblind
\author{\IEEEauthorblockN{\textbf{Anonymous Authors}}
	\IEEEauthorblockA{Anonymous Institutions\\
		e-mail: Anonymous Emails}}
\else
\author{\IEEEauthorblockN{\textbf{\paperAuthors}}
	\IEEEauthorblockA{$^1$Communication Networks Institute, $^2$Department of Computer Science, AI Group\\
		TU Dortmund University, 44227 Dortmund, Germany\\
		e-mail: \paperEmails}}
\fi

\maketitle

%% file: tex/arxiv.tex
\begin{tikzpicture}[remember picture, overlay]
\node[below=5mm of current page.north, text width=20cm,font=\sffamily\footnotesize,align=center] {Accepted for presentation in: 2020 IEEE International Conference on Communications (ICC)\vspace{0.3cm}\\\pdfcomment[color=yellow,icon=Note]{
@InProceedings\{Sliwa/etal/2020c,\\
  author    = \{Benjamin Sliwa and Nico Piatkowski and Christian Wietfeld\},\\
  title     = \{\{LIMITS\}: \{L\}ightweight machine learning for \{IoT\} systems with resource limitations\},\\
  booktitle = \{2020 IEEE International Conference on Communications (ICC)\},\\
  year      = \{2020\},\\
  address   = \{Dublin, Ireland\},\\
  month     = \{Jun\},\\
\}
}};
\node[above=5mm of current page.south, text width=15cm,font=\sffamily\footnotesize] {2020~IEEE. Personal use of this material is permitted. Permission from IEEE must be obtained for all other uses, including reprinting/republishing this material for advertising or promotional purposes, collecting new collected works for resale or redistribution to servers or lists, or reuse of any copyrighted component of this work in other works.};
\end{tikzpicture}

%% file: tex/abstract.tex
\begin{abstract}
	
%
%
Exploiting big data knowledge on small devices will pave the way for building truly cognitive \ac{IoT} systems.
%
%
Although machine learning has led to great advancements for \ac{IoT}-based data analytics, there remains a huge methodological gap for the deployment phase of trained machine learning models. For given resource-constrained platforms such as \acp{MCU}, model choice and parametrization are typically performed based on heuristics or analytical models. However, these approaches are only able to provide rough estimates of the required system resources as they do not consider the interplay of hardware, compiler-specific optimizations, and code dependencies.
%
%
In this paper, we present the novel open source framework \ac{LIMITS}, which applies a platform-in-the-loop approach explicitly considering the actual compilation toolchain of the target \ac{IoT} platform. \ac{LIMITS} focuses on high-level tasks such as experiment automation, platform-specific code generation, and sweet spot determination. The solid foundations of validated low-level model implementations are provided by the coupled well-established data analysis framework \ac{WEKA}.
%
We apply and validate \ac{LIMITS} in two case studies focusing on cellular data rate prediction and radio-based vehicle classification, where we compare different learning models and real world \ac{IoT} platforms with memory constraints from 16~kB to 4~MB and demonstrate its potential to catalyze the development of machine learning-enabled \ac{IoT} systems.   

\end{abstract}

%% file: tex/introduction.tex
\section{Introduction}

%
%
Ubiquitously deployed intelligent \ac{IoT} \cite{Zanella/etal/2014a} systems are expected to revolutionize smart city applications such as cognitive energy management \cite{Terroso-Saenz/etal/2019a} as well as smart logistics and intelligent traffic control in \acp{ITS} \cite{Sliwa/etal/2019b}.
%
%
Moreover, future 6G communication networks are envisioned to be massively exploiting data-driven methods \cite{Yang/etal/2019a} for network optimization in real time.

%
%
Although machine learning-enabled communication has been demonstrated to achieve massive performance improvements, e.g., through application of \emph{anticipatory} communication methods \cite{Bui/etal/2017a}, the deployment of such methods is not straightforward. 
State-of-the-art machine learning models are most often trained \emph{offline} in domain-specific languages---e.g., \texttt{python}, \texttt{R}, and \texttt{MATLAB}---however, resource-constrained \ac{IoT} platforms and \acp{MCU} typically operate on \texttt{C/C++} code. As automated processes for the transition from offline training to on-device prediction are missing, the deployment phase of the trained model can be rather cumbersome.
%
%
Moreover, for highly memory-constrained platforms---especially for ultra low power \acp{MCU}---even model choice and correct parametrization are challenging, as the trained models might easily exceed the capabilities of the target platform.
%
%
Pure analytical memory usage analyses of machine learning models are mostly \emph{optimistic} as they only consider the overall variable size of the model itself. External dependencies (e.g., mathematical libraries for the activation functions of \acp{ANN}) as well as the implementation of the logical data flow are not considered. However, the estimates can also be \emph{pessimistic} as they do not consider model size reduction techniques (e.g., pruning of \acp{CART}) and compiler-specific optimization.

\begin{figure}[b] 
	\centering
	\vspace{-0.5cm}
	
	\includegraphics[width=1\columnwidth]{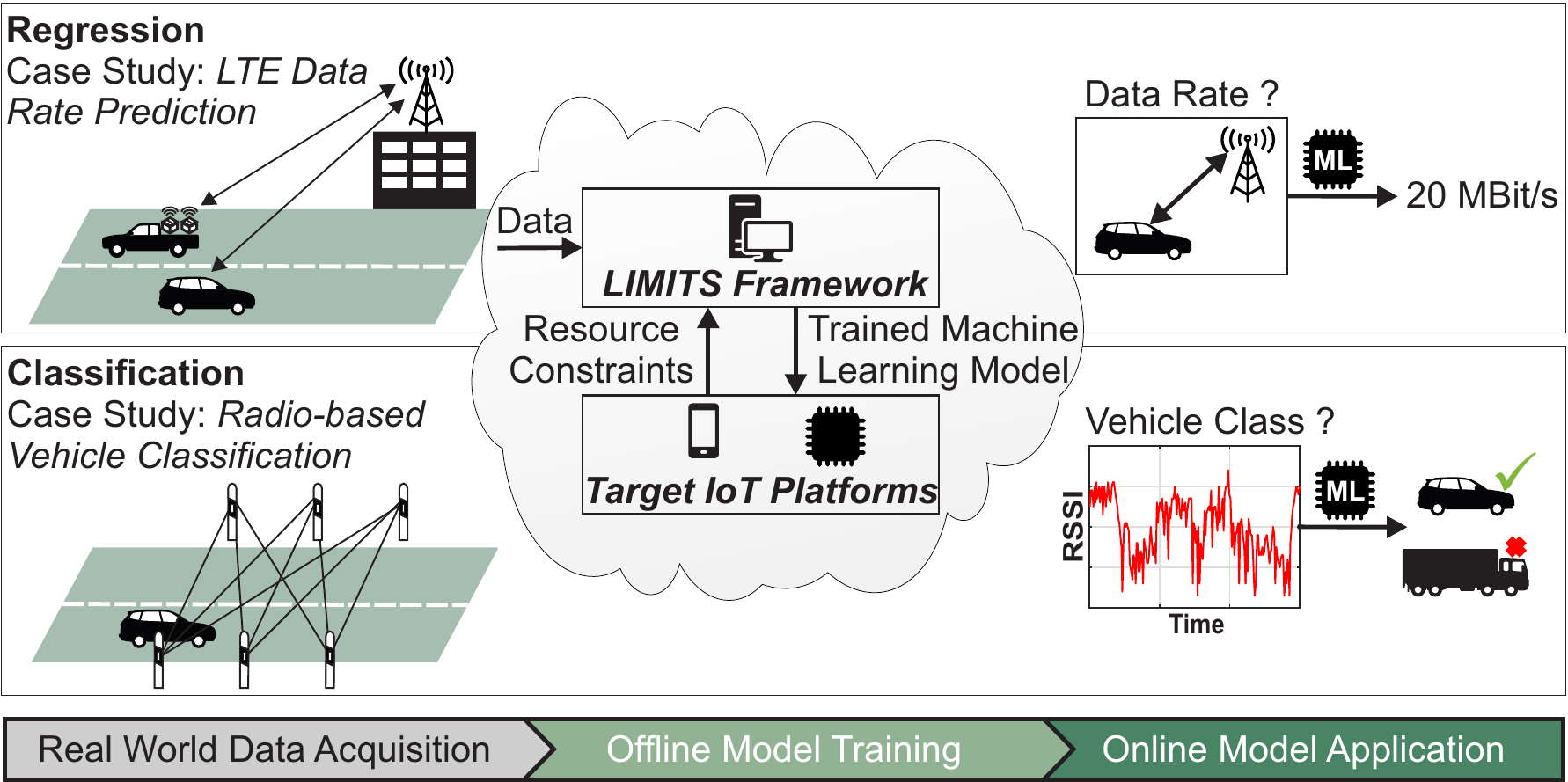}
	
	\caption{Example application scenarios: Models are trained offline and deployed to a target \ac{IoT} platform for performing online predictions.}
	\label{fig:scenario}
\end{figure}

%
%
In this paper, we present the novel open source framework \ac{LIMITS}, which aims to catalyze rapid prototyping of machine learning-enabled \ac{IoT} systems. \ac{LIMITS} provides an automated process for model comparison and parametrization with \emph{sweet spot} determination as well as generation of platform-specific \texttt{C/C++} code. The proposed framework implements a hybrid approach, where the low-level training of the machine learning models is performed using the well-known \ac{WEKA} framework. Considering the actual compilation toolchain of the target \ac{IoT} platform, \ac{LIMITS} is then able to perform a deep inspection of the trained model, including the actual memory consumption with all involved optimization steps. 
We illustrate typical data analytics tasks based on two case studies focusing on cellular data rate prediction and radio fingerprinting-based vehicle classification, both shown in Fig.~\ref{fig:scenario}.

%
%
%
%
%

%
%
The contributions provided by this paper can be summarized as follows:
\begin{itemize}
	\item Presentation of the novel \textbf{open source}\footnote{Source code is available at \url{https://github.com/BenSliwa/LIMITS}} \ac{LIMITS} framework for automating machine learning tasks (performance evaluation, model selection, model visualization) in the \ac{IoT} and wireless communication domains. 
	\item \textbf{Automatic generation of platform-specific \texttt{C/C++} code} for online application of the trained models on a given resource-constrained target platform and \textbf{platform-in-the-loop sweet spot} determination.
	\item \textbf{Proof-of-concept} study with real world data sets for \ac{LTE} uplink data rate prediction and radio-based vehicle classification and \textbf{deployment on different resource-constrained \ac{IoT}} platforms.
\end{itemize}
%
%
After discussing the related work in Sec.~\ref{sec:related_work}, we present the \ac{LIMITS} framework in Sec.~\ref{sec:approach}. The considered \ac{IoT} platforms and performance indicators are introduced in Sec.~\ref{sec:methods} and the case studies are discussed in Sec.~\ref{sec:results}.

%% file: tex/relatedWork.tex
\section{Related Work} \label{sec:related_work}

%
%
\textbf{Machine learning} has significantly stimulated research in cognitive optimization of wireless communication systems. Comprehensive introductions with a domain-specific perspective are provided by \cite{Liang/etal/2019a} and \cite{Ye/etal/2018a}. In this work, we focus on \emph{supervised} learning models for \emph{classification} and \emph{regression}---such as \ac{RF} \cite{Breiman/2001a}, \ac{M5} \cite{Quinlan/1992a}, \ac{SVM} \cite{Cortes/Vapnik/1995a}, and \ac{ANN} \cite{Goodfellow/etal/2016a}.
Recent advances in intelligent communication networks have demonstrated the potential benefits from using machine learning-based channel assessment for increasing the end-to-end data rate in vehicular cellular networks \cite{Sliwa/etal/2018b, Sliwa/Wietfeld/2019b}, for reinforcement learning-based opportunistic data transfer \cite{Sliwa/Wietfeld/2020a} as well as for optimizing power consumption of mobile \acp{UE} \cite{Sliwa/etal/2019a, Falkenberg/etal/2018a}, and \ac{DDNS} \cite{Sliwa/Wietfeld/2019c}. Although the considered regression case study in Sec.~\ref{sec:methods} focuses on client-based data rate prediction, first analyses have pointed out the potentials of cooperative prediction methods which might be realized in future 6G networks \cite{Sliwa/etal/2020b}.

%
%
Impressive results of machine learning---in particular with \emph{deep \acp{ANN}}---have been demonstrated, mostly in the computer vision and speech domains. However, in the communication networks domain, different principles apply, which influences the choice of analysis methods and models. Compared to aforementioned domains, the amount of training data is typically relatively low due to the highly application-centric nature of the analysis tasks. Data sets are usually acquired manually based on real world experiments. 
As a consequence, deep learning methods are often outperformed by simpler approaches, e.g., \ac{CART}-based models, which require less data to reach a satisfying performance level \cite{Sliwa/Wietfeld/2019b}. A positive side effect---which is exploited for sweet spot determination in Sec.~\ref{sec:sweet_spot}---of the small data sets is the comparably short training time, which allows to train and compare a multitude of different models and parameterizations for a given task.

%
%
\textbf{Resource-constrained \ac{IoT} systems} have raised a keen interest of the research community due to their sensing and communication capabilities as well as their cost efficiency \cite{Park/etal/2016a, Yao/etal/2018a}.
%
%
In \cite{Masoudinejad/etal/2018a}, the authors compare the classification performance of different models for a positioning task in an industrial environment. Although a memory-constrained target platform is considered, the memory consumption of the machine learning models is only derived analytically based on model complexity considerations. However, as investigated in Sec.~\ref{sec:results}, there is a significant gap between analytical estimations and the actual platform-specific resource occupation.

As an alternative approach for deploying generic machine learning models on \ac{IoT} platforms, other approaches aim to optimize the memory efficiency of the algorithms themselves. E.g., model compression techniques \cite{Bucila/etal/2006a,Piatkowski/etal/2013a} stem from the machine learning community and do not take the actual hardware into account. 
%
%
In \cite{Kumar/etal/2017a}, the \emph{Bonsai} model is presented which consists of a sparse tree model that is learned in a low-dimensional feature space through projection. The model is deployed to a \texttt{ATmega328P} platform---which is also considered in this paper---with 2~KB \ac{RAM} and 32~KB program memory. In comparison to cloud-based offloading (e.g., as proposed by \cite{Shukla/Munir/2017a}) the local execution consumes 47-497 times less energy.
%
%
%
%
Another optimization approach is the avoidance of floating point arithmetic. In \cite{Piatkowski/etal/2016a}, the authors propose integer undirected models for performing probabilistic inference on resource-constrained systems.

\textbf{Data analysis tools}: In addition to well-established mathematical tool such as \texttt{MATLAB} and \texttt{R}, a wide range of different machine learning tools has emerged.
%
%
RapidMiner \cite{Hofmann/Klinkenberg/2013a} is a commercial and graphical data analysis framework focusing on business analytics. Although a free version is provided for academic usage, it is limited in the number of data values. \ac{KNIME} \cite{Berthold/etal/2009a} provides similar functions and an open source licensing model.
%
%
\ac{WEKA} \cite{Hall/etal/2009a} is a \texttt{Java}-based open source framework for machine learning with limited graphical features but a rich \ac{CLI}, which makes it a powerful tool for automation.

%
%
Recently, \texttt{python}-based frameworks focusing on deep learning such as Scicit-learn \cite{Pedregosa/etal/2011a}, PyTorch \cite{Paszke/etal/2017a}, and TensorFlow \cite{Abadi/etal/2016a} enjoy a great popularity. Although the latter toolkit has made initial attempts to support data analysis on embedded devices with \emph{TensorFlow Lite}, the available models and supported platforms are limited and focus on typical \ac{ANN} tasks such as image recognition.

%% file: tex/solution_approach.tex
\section{Machine Learning with \framework} \label{sec:approach}

In this section, we introduce the proposed \framework framework and highlight different capabilities focusing on automating data analytics and deployment of trained models to \ac{IoT} platforms.
%
%
\begin{figure}[] 
	\centering
	
	\includegraphics[width=1\columnwidth]{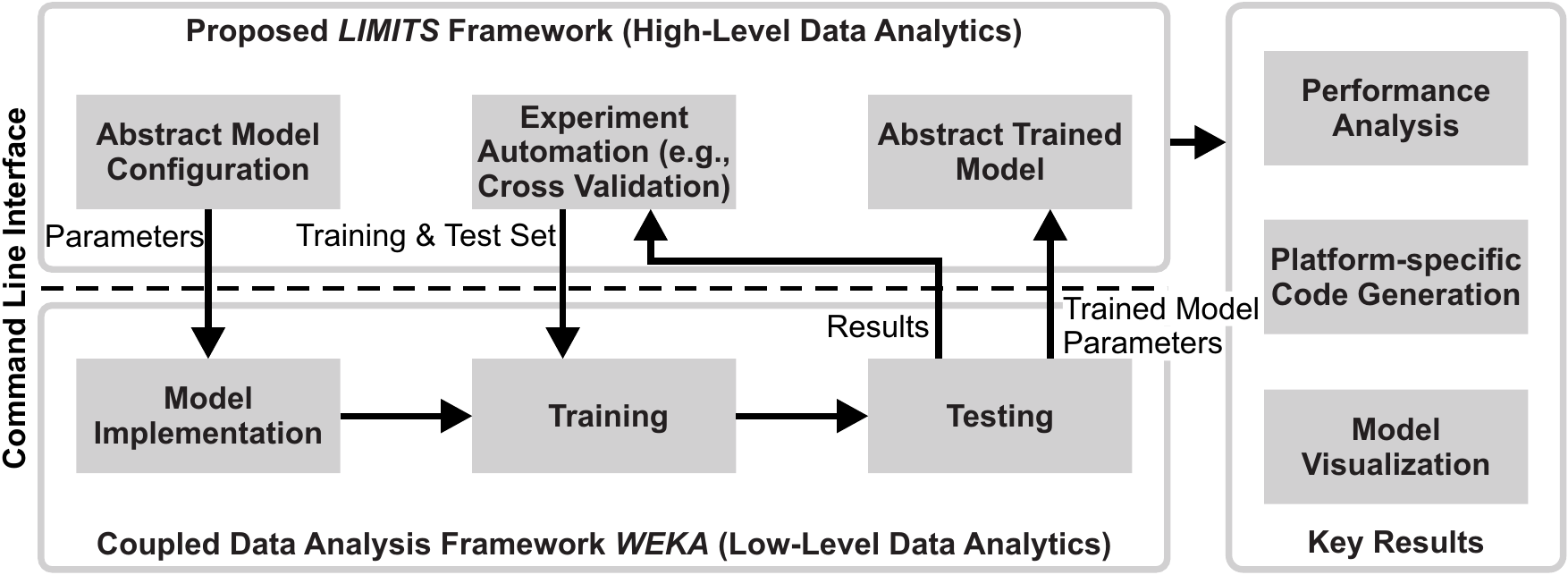}
	
	\caption{Overall system architecture of the proposed \ac{LIMITS} data analysis framework.}
	\vspace{-0.5cm}
	\label{fig:architecture}
\end{figure}
The general process for performing machine learning-based data analysis with \ac{LIMITS} is illustrated in Fig.~\ref{fig:architecture}. 
%
%
\ac{LIMITS} provides a convenient \texttt{python}-based interface for performing high-level machine learning tasks, ready to analyze the performance of multiple models or configurations. It uses an abstract model configuration, for which the parameters are close to the analytical model descriptions. 
%
%
The actual low-level machine learning processes are performed by \ac{WEKA}, which provides a solid foundation of validated model implementations. 
%
%
Based on the reported result parametrization of the trained model (e.g., weight matrices for \acp{ANN}, abstract trees for \acp{CART}), \ac{LIMITS} derives an abstract intermediate representation of the trained model, which can then be used to generate platform-specific code.

%
%
As \ac{LIMITS} only considers abstract models and the whole interaction with the coupled \ac{WEKA} is performed through dedicated interfaces, it is highly extensible and can be extended to support additional low-level frameworks in the future.

%
%
	\begin{lstlisting}[caption={Example for CLI-based machine learning for rapid result analysis.},captionpos=b,label=lst:cli]
	$ ./cli.py -r ../examples/mnoA.csv -m rf,m5,ann
	
	r2               mae              rmse
	0.78+/-0.01      2.95+/-0.24      4.01+/-0.21
	0.78+/-0.02      2.76+/-0.12      3.99+/-0.18
	0.79+/-0.01      3.34+/-0.33      4.35+/-0.32\end{lstlisting}
Lst.~\ref{lst:cli} shows an example usage of the \ac{CLI} where three different models (\ac{RF}, \ac{M5}, and \ac{ANN}) are applied to 
perform a regression task on the data set \texttt{mnoA.csv}. It can be seen that the \ac{CLI} provides a lightweight interface for rapidly assessing the performance of different models. 
More detailed model configurations and parameter definitions can be performed based on the integrated \texttt{python} interface.

\subsection{Machine Learning Models} \label{sec:models}

In its current version, the proposed \ac{LIMITS} framework supports data analysis and code generation for different well-known \emph{supervised} machine learning models.

%
%
\textbf{\acfp{ANN}} \cite{Goodfellow/etal/2016a} consist of different layers of interconnected nodes which are referred to as \emph{neurons} in analogy to the cognitive systems of living creatures.
%
%
Recently, \emph{Deep Neural Networks}, which are \acp{ANN} with more than one hidden layer, have received great attention due to their superior performance in tasks such as image classification.
Those models can be implemented as a series of matrix vector multiplications followed by component-wise neuron activation functions. The resulting memory occupation is mainly related to the weight matrices and threshold vectors. For a fully connected \ac{ANN} with a layer configuration $L$ of $N$ layers, the resulting memory usage $\text{M}_{\text{ANN}}$ is estimated as
\begin{equation} \label{eq:ann_resources}
	\text{M}_{\text{ANN}} = \sum\limits_{i=2}^{N} L_{i} (1 + L_{i-1}) 
\end{equation}

%
%
\textbf{\acf{SVM}} \cite{Cortes/Vapnik/1995a} learn a hyperplane that separates real-valued data points in a $d$ dimensional hyperspace by minimizing a specific objective function. $d$ corresponds to the dimensionality of the provided feature vector.
%
%
\ac{LIMITS} currently provides code generation for linear L2/L2 \acf{SVM}. For multi-class problems with $n$ classes, the \emph{one-vs-one} strategy is applied and the memory usage $M_{\text{SVM}}$ can be estimated as 
\begin{equation}
	M_{\text{SVM}} = \frac{n(n-1)}{2} d
\end{equation}
%
%
%

%
%
\textbf{\acf{M5}} \cite{Quinlan/1992a} is a regression tree method where each of the leaves contains a linear regression model. The resulting memory consumption of tree-based methods cannot be easily calculated proactively, as the tree structure is derived within the training process and its optimization steps. Although it is possible to derive worst case estimations for the tree size, the resulting value range is not meaningful enough for the considered \ac{MCU} platforms which require fine-grained model size estimations. \ac{M5} can be applied for regression but not for classification.

\begin{figure}[] 
	\centering
	
	\includegraphics[width=1\columnwidth]{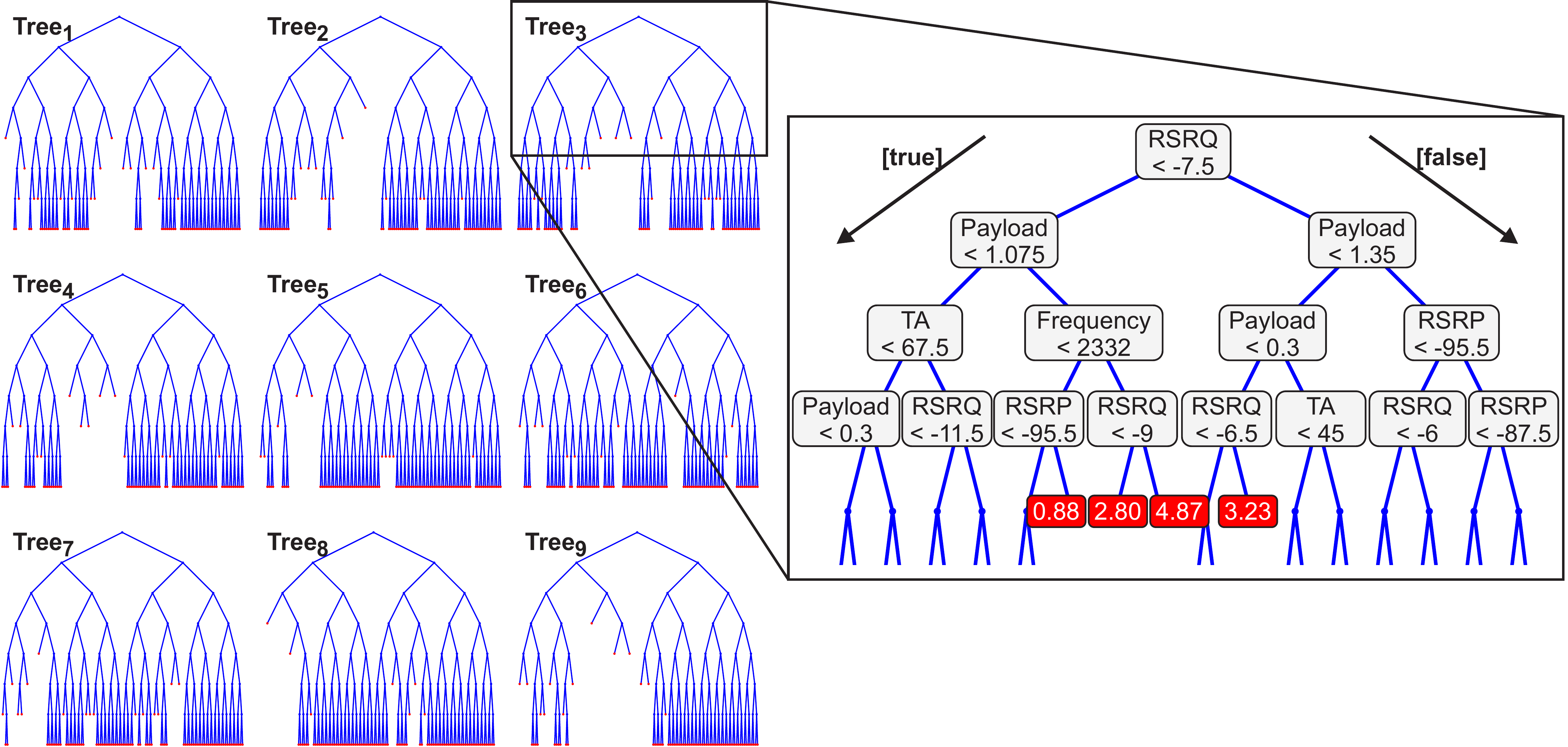}
	
	\vspace{-0.2cm}
	\caption{Example model visualization for a trained \acf{RF} with nine trees and maximum depth 7. The overlay shows an excerpt of the actual decision making within one of the trees.}
	\vspace{-0.1cm}
	\label{fig:rf_vis}
\end{figure}

%
%
\textbf{\acfp{RF}} \cite{Breiman/2001a} in an ensemble method based on multiple randomized \ac{CART} trees where each tree is learned from a random sub-set of the training data. The leaves contain the prediction results and the final result is obtained by averaging over all tree results. An example visualization of a trained \ac{RF} is shown in Fig.~\ref{fig:rf_vis}. \acp{RF} are frequently used for network quality prediction (e.g., \cite{Sliwa/Wietfeld/2019b, Jomrich/etal/2018a, Samba/etal/2017a} and often outperform methods which require a larger amount of training data for reaching a similar level of prediction performance.
%
%

\subsection{Automation of Data Analytics Tasks}

%
%
\begin{figure}[] 
	\centering
	
	\includegraphics[width=1\columnwidth]{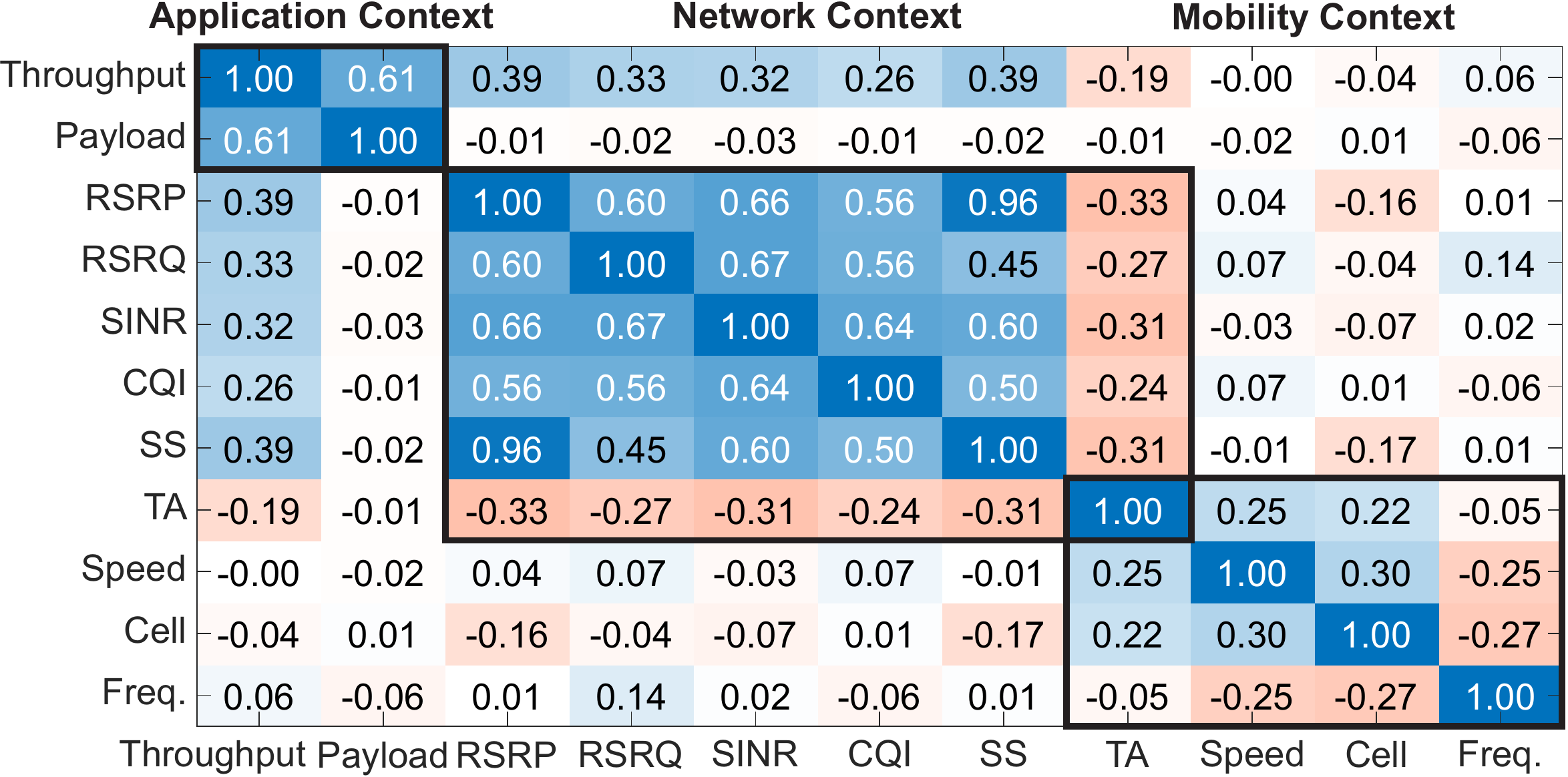}
	
	\caption{Correlation-based feature analysis of the Android-based cellular network measurements. The different context classes can be clearly identified.}
	\vspace{-0.5cm}
	\label{fig:feature_correlation}
\end{figure}

\ac{LIMITS} supports the automatic execution of different data analysis tasks:
\begin{itemize}
	\item \textbf{Correlation analysis} is an intuitive way of feature selection for removing redundancies within a data set \cite{Apajalahti/etal/2018a}. Fig.~\ref{fig:feature_correlation} shows an example data set of Android-based \ac{LTE} performance indicator measurements (further details about the methodological aspects on the features are discussed in Sec.~\ref{sec:methods}). It can be seen that \ac{RSRP} and \ac{SS} have a high cross-correlation. In fact, the  \ac{SS} reported by the Android operating system represents the \ac{ASU} which is derived as $\text{ASU} = \text{RSRP} + 140$. Such redundancies should be removed from the data set in order to optimize the information gain or whenever machine learning methods assume independent features. 	
	\item \textbf{Cross validation} considers multiple training and test scenarios from one data set in order to achieve deep insights into a model's mean performance as well as its standard deviation. Relying on a single train/test split of the data could deliver highly over- or underconfident estimates of a model's performance. As \framework focuses on real world model deployments, a more robust assessment approach is therefore applied.
	\item \textbf{Experiments} are used to compare the performance of multiple models/parametrizations on a single data set. 
	\item \textbf{Multi experiment} analyze the performance of a single model on $N$ data sets. This method is applied to analyze if a machine learning model generalizes well on different data sets (e.g., different evaluation tracks, network data for multiple \acp{MNO} \cite{Sliwa/Wietfeld/2019b}). The results are $N\text{x}N$ matrices for each performance indicator.
	\item \textbf{Feature importance} analysis based on model-specific indicators, e.g., \ac{MDI} \cite{Louppe/etal/2013a}. 
\end{itemize}

\subsection{Platform-in-the-Loop Model Selection} \label{sec:sweet_spot}

The proposed platform-in-the-loop approach allows us to determine the Pareto optimal configuration---the \emph{sweet spot}---with respect to quality and resource allocation of each model for a given platform. More precisely, the goal is to find the model parametrization with the highest prediction accuracy, which just fulfills the memory requirements of the target platform.

%
%
\begin{figure}[] 
	\centering

	\includegraphics[width=1\columnwidth]{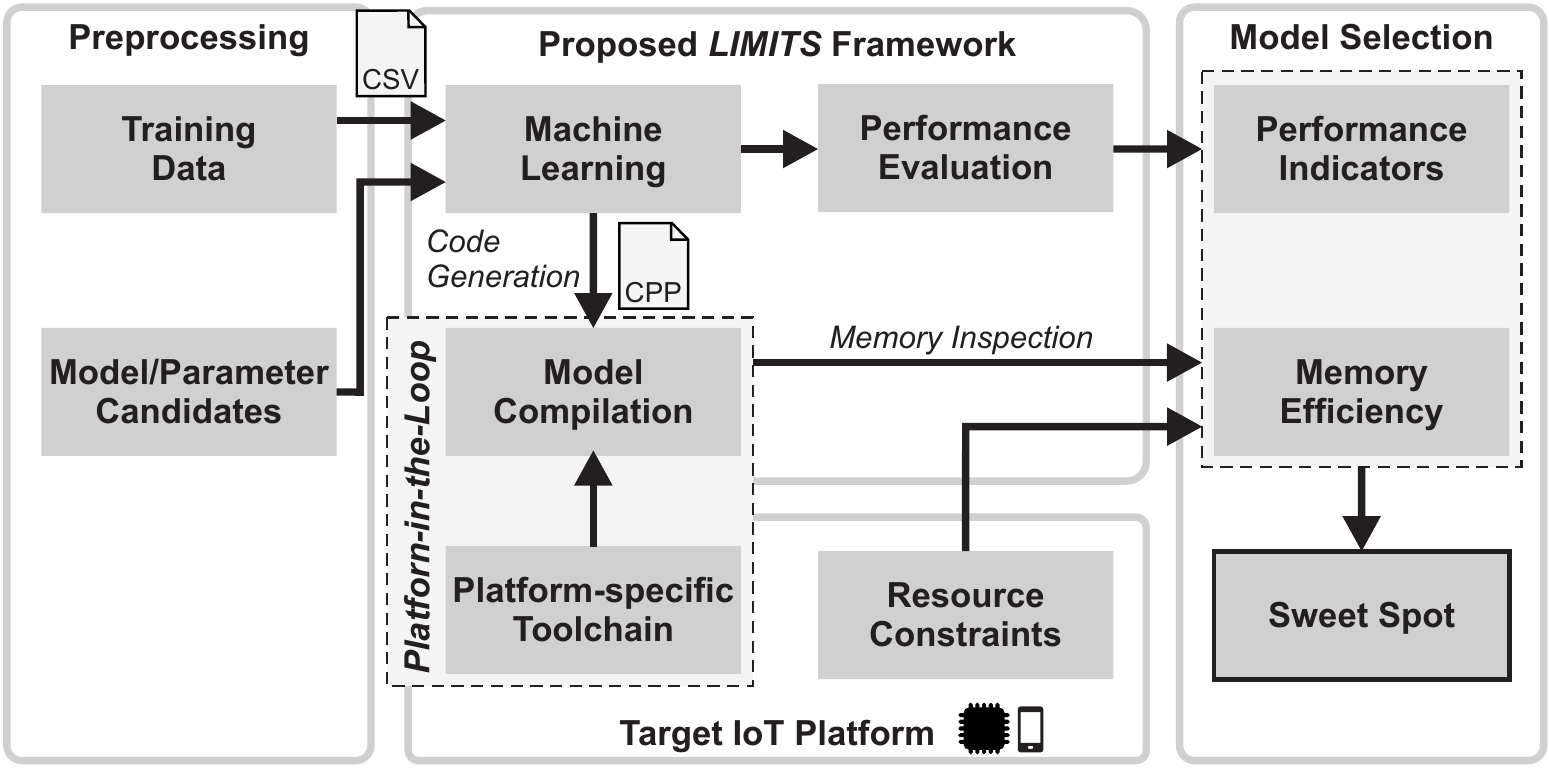}
	
	\caption{Platform-in-the-loop model selection process which is performed with respect to the  platform-specific compilation toolchain in order to determine the \emph{sweet spot} of the model.}
	\vspace{-0.5cm}
	\label{fig:modelSelection}
\end{figure}
An overview about the provided automated model selection process is shown in Fig.~\ref{fig:modelSelection}. For each \emph{candidate}, a model is trained on the supplied training data and the corresponding \texttt{C/C++} code for the model is generated automatically. The latter is compiled with the toolchain of the target platform, which allows to inspect the real resulting memory occupation. Finally, the sweet spot is determined based on the achieved model performance and its resource efficiency.

%% file: tex/methodology.tex
\section{Methodology} \label{sec:methods}

In the following, we discuss the case studies, performance indicators, and target \ac{IoT} platforms.

\subsection{Case Studies and Data Sets}

As the major contribution of this work is of methodological nature, we utilize existing data sets, which have been acquired in previous work.

%
%
\textbf{Regression} is considered with a case study focusing on client-based \ac{LTE} uplink data rate prediction in vehicular scenarios \cite{Sliwa/Wietfeld/2019b}. Measurements for the context indicators \ac{RSRP}, \ac{RSRQ}, \ac{SINR}, \ac{CQI}, \ac{TA}, velocity, cell id, carrier frequency, and payload size are utilized by the \ac{UE} to forecast the achievable data rate for each to be performed \ac{TCP} data transfer. The considered data set consists of 3907 transmissions, which were performed in the public cellular network and in four different scenarios (urban, suburban, highway, campus).

%
%
\textbf{Classification} is considered with a radio-based vehicle classification system \cite{Sliwa/etal/2018e}.
Hereby, an installation of six communicating IEEE 802.15.4 nodes is installed with three nodes on each of the road sides (see the illustration in Fig.~\ref{fig:scenario}). The \ac{RSSI} level of the resulting nine different radio links $\Phi_{i}$ is monitored continuously.
If a vehicle passes the installation, the resulting attenuation pattern results in a \emph{radio fingerprint}, which is characteristic for the type of the vehicle. Seven different vehicle classes are considered: \emph{Passenger car, Passenger car with trailer, Van, Truck, Truck with Trailer, Semitruck, Bus}. The considered data set consists of 2605 time series traces.

\subsection{Performance Indicators}

%
%
The performance of the data rate prediction models is assessed by means of the \emph{coefficient of determination} $R^{2}$ (a.k.a. amount of explained variance), which is a statistical metric and allows to compare the achieved results with other related approaches that consider the same indicators \cite{Jomrich/etal/2018a, Samba/etal/2017a}. It is calculated as
%
%
\begin{eqnarray}
	R^{2} = 1- \frac{\sum_{i=1}^{N}\left(\tilde{y}_{i} - y_{i} \right)^{2}}{\sum_{i=1}^{N}\left(\bar{y} - y_{i} \right)^{2}}
\end{eqnarray}
with $\tilde{y}_{i}$ being the current prediction, $y_{i}$ being the current measurement, and $\bar{y}$ being the mean measurement value. Further performance metrics for the validation of the code generator are \ac{MAE} and \ac{RMSE}.

%
%
For the vehicle classification task, we mainly focus on analyzing \emph{accuracy} (which is also referred to as \emph{classification success ratio} in related work). For a given data set $\mathcal{D}$, a model $f$ is trained to make predictions on unlabeled data $\mathbf{x}$ such that $\tilde{y} = f(\mathbf{x})$. The accuracy \text{ACC} is then derived as
%
%
\begin{equation}
	\text{ACC}(f;\mathcal{D}) = \frac{1}{|\mathcal{D}|} \sum_{(y, \mathbf{x})\in\mathcal{D}}1_{\left\lbrace y=f(\mathbf{x})\right\rbrace}
\end{equation}
with $|\mathcal{D}|$ being the cardinality of $\mathcal{D}$ and $1_{\left\lbrace y=f(\mathbf{x})\right\rbrace}$ being the indicator function that only evaluates to 1 if $f(\mathbf{x})$ outputs the correct class $y$ and is 0 otherwise.
For the validation of the code generator, we further consider precision, recall and $\text{F}_{1}$-score.

%
%

\subsection{Target \ac{IoT} Platforms for the Performance Evaluation}

For the performance evaluation in Sec.~\ref{sec:results}, we evaluate three popular target \ac{IoT} platforms with different resource requirements and computation capabilities.

%
%
\textbf{\msp} (Model G2553) is a 16~Bit ultra low power microcontroller with 16~MHz, 16.35~kB program memory and 512~Byte \ac{RAM}, which is programmed with \texttt{Code Composer Studio}. The compiler is configured to apply memory-centric optimization with disabled \emph{loop unrolling} features.

%
%
\textbf{\atmega} (Model ATmega328P) is a 16~MHz \ac{MCU} with 32~kB program memory 2~kB \ac{RAM}. Compilation and deployment are performed with the popular \texttt{Arduino} toolkit.

%
%
\textbf{\esp} (Model ESP-WROOM-32) is a 32~Bit \ac{MCU} with a 240~MHz dual core, 4~MB program memory and 532~kB \ac{RAM} with integrated bluetooth and WiFi support. Compilation and deployment tools are provided by \texttt{\ac{ESP-IDF}}.

%% file: tex/results.tex
\section{Results} \label{sec:results}

%
%
In this section, models for a given \ac{IoT} platform are selected and the generated model implementations are validated against the \ac{WEKA} results. All data analysis evaluations are 10-fold cross validated.

\subsection{Platform-specific Model Selection}

%
%
Sweet spots are identified by optimizing the model specific hyper-parameters on each platform.

%
%
For the \ac{ANN}, we analyze the impact of different amounts of hidden layers and the number of nodes on each hidden layer. 
%
%
\begin{figure*}[]
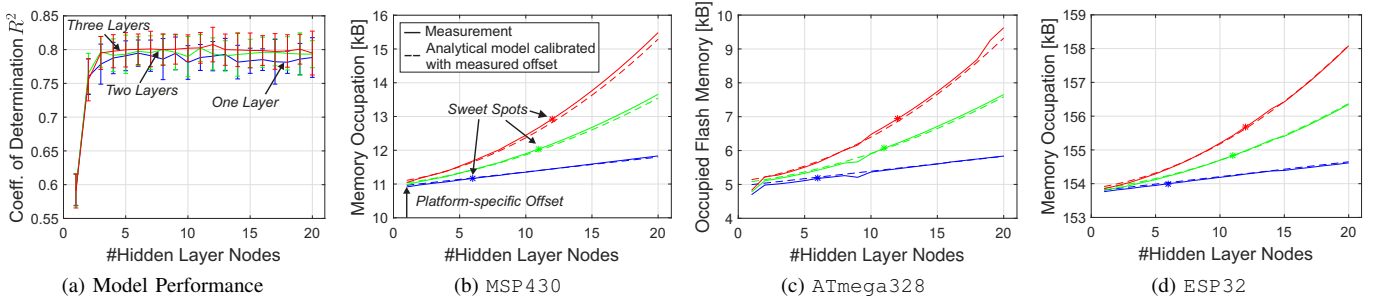
 
	\centering
	
	\subfig{fig/ann_reg_acc}{\sfw}{Model Performance}%
	\subfig{fig/ann_reg_msp}{\sfw}{\msp}%
	\subfig{fig/ann_reg_atmega}{\sfw}{\atmega}%
	\subfig{fig/ann_reg_esp}{\sfw}{\esp}%
	
	\caption{Sweet spot determination for the data rate prediction data set with \ac{ANN} and different values for number of hidden layers and number of nodes per hidden layer. The error bars show the standard deviations of the 10-fold cross validation.}
	\vspace{-0.5cm}
	\label{fig:ann_reg_parameters}
\end{figure*}					
Fig.~\ref{fig:ann_reg_parameters} shows the resulting coefficient of determination $R^2$ and the occupied program memory resources for all platforms on the data rate prediction data set.
%
%
Although it can be seen that the analytical model with 4 Byte \texttt{float} data types (see Eq.~\ref{eq:ann_resources}) provides a good estimate for the real memory occupation, it contains a \emph{platform-specific offset} which is unknown if only the analytical model is considered.
%
%
For the ESP32, it needs to be denoted that the memory footprint can be reduced by disabling communication capabilities from the compilation process. However, as we aim to mimic typical application scenarios, we kept the default configuration of the \ac{MCU}.

%
%
\begin{figure*}[]
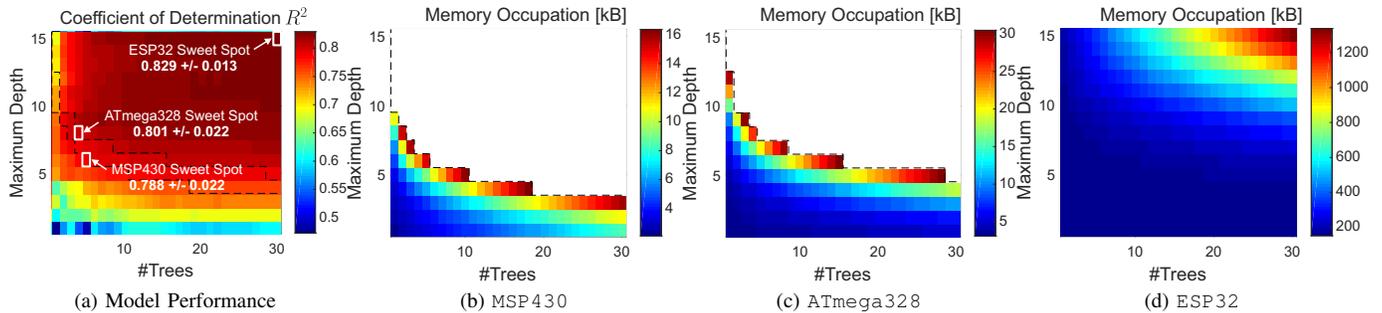
 
	\centering
	
	\subfigh{fig/rf_reg_acc}{\sfw}{Model Performance}%
	\subfigh{fig/rf_reg_msp}{\sfw}{\msp}%
	\subfigh{fig/rf_reg_atmega}{\sfw}{\atmega}%
	\subfigh{fig/rf_reg_esp}{\sfw}{\esp}%
	
	\caption{Sweet spot determination for the data rate prediction data set with \ac{RF} and different values for number of random trees and maximum depth.}
	\vspace{-0.5cm}
	\label{fig:rf_reg_parameters}
\end{figure*}					
For the \ac{RF} model, we vary the number of random trees and the maximum tree depth. Fig.~\ref{fig:rf_reg_parameters} shows the results for the data rate prediction data set. While the \esp is able to consider the whole parameter space, \atmega and \msp are significantly impacted by memory limitations.


For each model, the results of the platform-specific sweet spot parametrizations are summarized in Tab.~\ref{tab:regression} for the regression task and in Tab.~\ref{tab:classification} for the classification task with sweet spot parameters for \ac{ANN} and \ac{RF}.

%
%
\input{tex/tables/regression}

%
%
\input{tex/tables/classification}

%
%
%

\subsection{Runtime Complexity}
\begin{figure}[] 
	\centering
	
	\includegraphics[width=1\columnwidth]{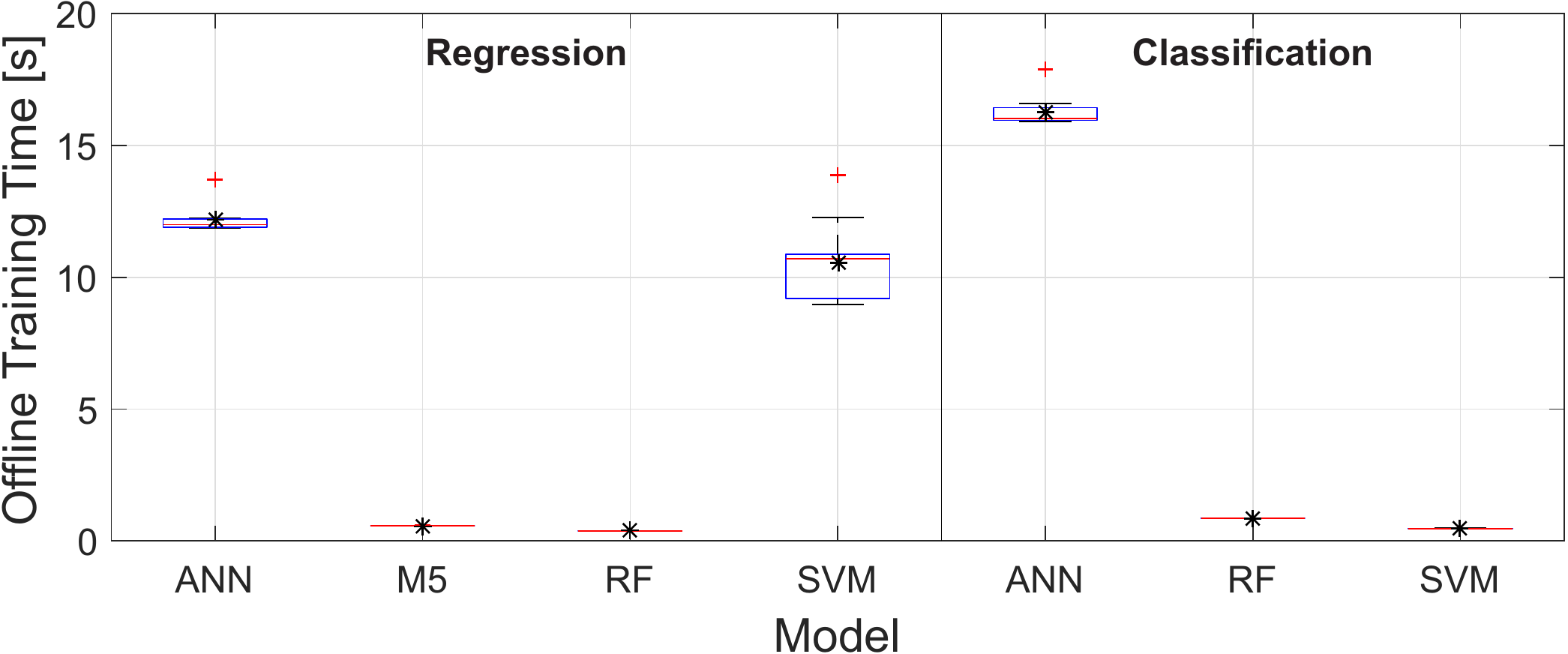}
	
	\caption{Runtime of training and testing the machine learning models (offline).}
	\label{fig:times_offline}
	
\end{figure}

%
%
Enumerating the resource requirements of the full 
hyper-parameter space (as shown in Fig.~\ref{fig:rf_reg_parameters}) comes with an undeniable computational overhead. Models which allow for fast training are thus especially well suited for our proposed system. The training times of considered machine learning models are shown in Fig.~\ref{fig:times_offline}: M5 and random forests clearly outperform ANN and SVM training, which implies that sweet spots of these method can be identified much faster in practice. Stochastic gradient methods are known to suffer from sub-linear convergence, which explains the inferior runtime of ANN training. SVM, on the other hand, is trained with Platt's sequential minimal optimization. While this algorithm can be very fast occasionally, it's worst-case behavior is quadratic in the number of data points---this is likely to happen whenever almost all data points will be support vectors of the underlying model.

%
%
%
%
%

%
%
While training time can be mitigated by using strong computational resources, prediction time of a model on the resource-constrained device can render a task impractical. Thus, we investigate how the program code that is generated by \ac{LIMITS} performs on each platform. 
In the following, the best model is deployed for each target \ac{IoT} platform and the execution time per single prediction is determined over 1000 online predictions.
%
%
\begin{figure}[] 
	\centering
	
	\includegraphics[width=1\columnwidth]{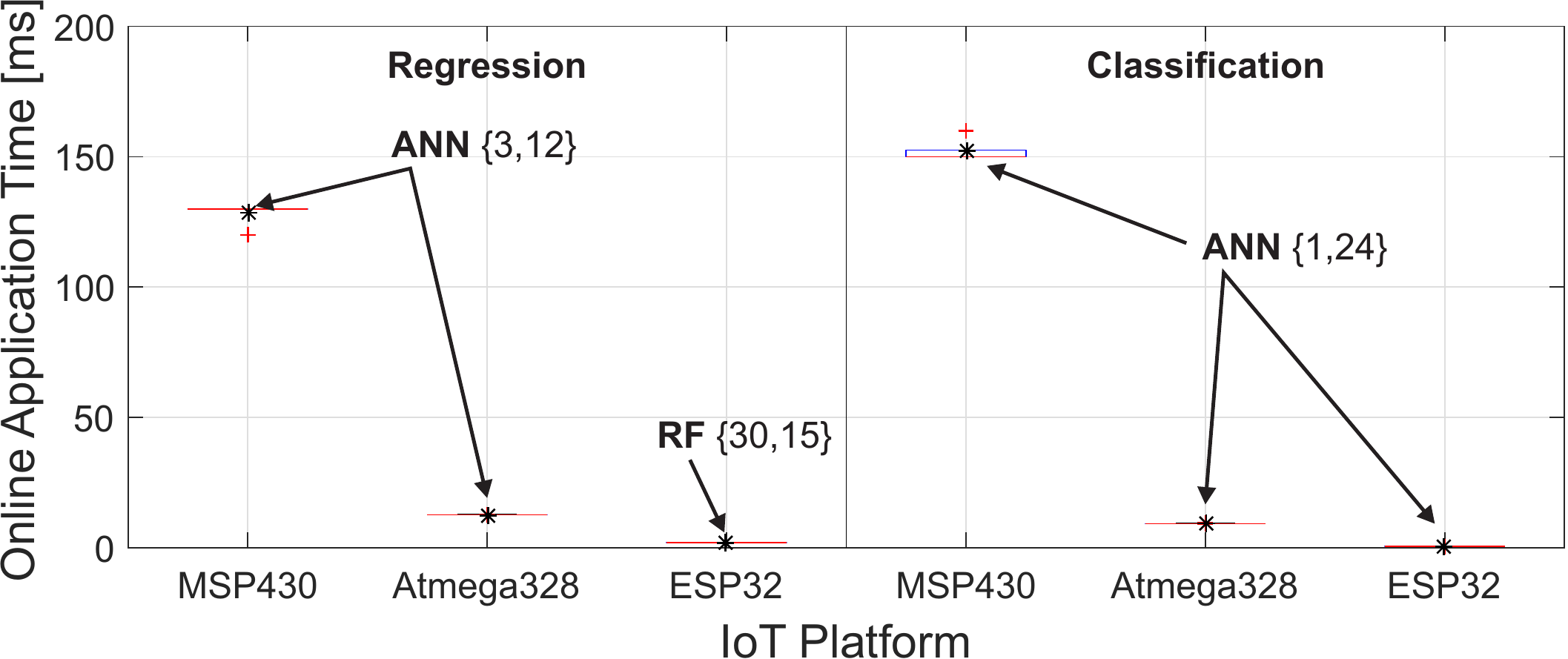}
	
	\caption{Measured online execution time per prediction of the sweet spot models deployed on the considered target platforms. \emph{\ac{ANN}}~\{\#Hidden layers, \#Neurons on hidden layers\}, \emph{\ac{RF}}~\{\#Trees, \#Max. depth\}.}
	\label{fig:times_online}
\end{figure}
Fig.~\ref{fig:times_online} shows the resulting measurement values for the regression and classification tasks. 

Each test platform achieves prediction rates between $8 \text{Hz}$ and $100 \text{Hz}$---sufficient for our example applications. Nevertheless, the \msp which has by far the lowest energy requirements is outperformed by a large margin by both, the \atmega and the \esp platforms.

\subsection{Code Generator Validation}

In order to validate the code generator implementation, we replay all measurements with the generated \texttt{C/C++} models and compare the statistical properties to the \ac{WEKA} results.
The results are shown in Tab.~\ref{tab:validation_reg} for the regression models and in Tab.~\ref{tab:validation_class} for the classification models.
%
%
\input{tex/tables/validation_reg}

%
%
\input{tex/tables/validation_class}
%
%
%
It can be seen that the generated models achieve an accurate match with the ground truth provided by \ac{WEKA}. However, minor deviations occur as the the latter exposes rounded parameters for some of the machine learning models.

%% file: tex/tables/regression.tex
\newcolumntype{Y}{>{\centering\arraybackslash}X}
\newcommand\mr[1]{\multirow{2}{1cm}{#1}}

\begin{table}[ht]
	\centering
	\caption{Regression Model Performance and Program Memory Occupation for Different \acs{IoT} Platforms}
	\vspace{-0.3cm}
		
	\begin{tabularx}{\columnwidth}{l c*{6}{Y}}
		\toprule
		\textbf{Model} & \multicolumn{2}{c}{\textbf{\msp}} & \multicolumn{2}{c}{\textbf{\atmega}} & \multicolumn{2}{c}{\textbf{\esp}} \\
		\cmidrule(lr){2-3} \cmidrule(l){4-5} \cmidrule(l){6-7}
		& $\mathbf{R^2}$ & \textbf{Memory} & $\mathbf{R^2}$ & \textbf{Memory} & $\mathbf{R^2}$ & \textbf{Memory}\\
		\midrule
		
		%
		%
		{\textbf{\acs{ANN}}} & $\textbf{0.807}$ & {\textbf{12.62~kB}} & $\textbf{0.807}$ & {\textbf{6.77~kB}} & $0.807$ & {152.03~kB} \\
		$\{\text{H},\text{N}\}$ & $\pm0.025$ & $\{3,12\}$ & $\pm0.025$ & $\{3,12\}$ &  $\pm0.025$ & $\{3,12\}$\\
		 
		%
		%
		\multirow{2}{0.5cm}{\textbf{\acs{M5}}} & $0.772$ & \mr{5.15~kB} & $0.772$ & \mr{3.7~kB} & $0.772$ & \mr{149.80~kB}\\
		& $\pm0.03$ & & $\pm0.03$ & & $\pm0.03$ &\\
		
		%
		%
		{\textbf{\acs{RF}}} & $0.788$ & {14.59~kB} & $0.801$ & {24.76~kB} & $\textbf{0.829}$ & {\textbf{1307~kB}} \\
		$\{\text{T},\text{D}\}$ & $\pm0.022$ & $\{5,6\}$ & $\pm0.022$ & $\{4,8\}$ & $\pm0.013$ & $\{30,15\}$ \\
		
		%
		%
		\multirow{2}{0.5cm}{\textbf{\acs{SVM}}} & $0.551$ & \mr{3.72~kB} & $0.551$ & \mr{7.88~kB} & $0.551$ & \mr{148.70~kB}\\
		& $\pm0.03$ & & $\pm0.03$ & & $\pm0.03$ &\\
		
		\bottomrule
	\end{tabularx}
	
	\vspace{0.1cm}
	\emph{H}: \#Hidden layers, \emph{N}: \#Neurons on hidden layer, \emph{T}: \#Trees, \emph{D}: Max. depth
	\label{tab:regression}
\end{table}

%% file: tex/tables/classification.tex
\newcolumntype{Y}{>{\centering\arraybackslash}X}
\renewcommand\mr[1]{\multirow{2}{1cm}{#1}}

\begin{table}[ht]
	\centering
	\caption{Classification Model Performance and Program Memory Occupation for Different \acs{IoT} Platforms}
	\vspace{-0.3cm}
	
	\begin{tabularx}{\columnwidth}{l c*{6}{Y}}
		\toprule
		\textbf{Model} & \multicolumn{2}{c}{\textbf{\msp}} & \multicolumn{2}{c}{\textbf{\atmega}} & \multicolumn{2}{c}{\textbf{\esp}} \\
		\cmidrule(lr){2-3} \cmidrule(l){4-5} \cmidrule(l){6-7}
		& \textbf{Accuracy} & \textbf{Memory} & \textbf{Accuracy} & \textbf{Memory} & \textbf{Accuracy} & \textbf{Memory}\\
		\midrule
		
		%
		%
		{\textbf{\acs{ANN}}}  & $\textbf{93.79}$ & {\textbf{13.24~kB}} & $\textbf{93.79}$ & {\textbf{6.51~kB}} & $\textbf{93.79}$ & {\textbf{152.4~kB}}\\
		$\{\text{H},\text{N}\}$ & $\pm1.15$ & $\{1,24\}$ & $\pm1.15$ & $\{1,24\}$  & $\pm1.15$ & $\{1,24\}$\\

		%
		%
		{\textbf{\acs{RF}}} & $93.05$ & {16.12~kB} & $93.48$ & {28.54~kB} & $93.67$ & {256.71~kB} \\
		$\{\text{T},\text{D}\}$ & $\pm0.87$ & $\{5,5\}$ & $\pm1.32$ & $\{5,13\}$ & $\pm1.37$ & $\{12,13\}$\\

		%
		%
		\multirow{2}{0.5cm}{\textbf{\acs{SVM}}} & $92.48$ & \mr{5.45~kB} & $92.48$ & \mr{9.22~kB} & $92.48$ & \mr{151.04~kB}\\
		& $\pm0.62$ & & $\pm0.62$ & & $\pm0.62$ &\\
		
		\bottomrule
	\end{tabularx}
	
	\vspace{0.1cm}
	\emph{H}: \#Hidden layers, \emph{N}: \#Neurons on hidden layer, \emph{T}: \#Trees, \emph{D}: Max. depth
	\label{tab:classification}
\end{table}

%% file: tex/tables/validation_reg.tex
\newcolumntype{Y}{>{\centering\arraybackslash}X}

\begin{table}[ht]
	\centering
	\caption{Validation of the Code Generator for Regression Models}
	\vspace{-0.3cm}
	
	\begin{tabularx}{\columnwidth}{l c*{6}{Y}}
		\toprule
		\textbf{Model} & \multicolumn{3}{c}{\textbf{\acs{WEKA}}} & \multicolumn{3}{c}{\textbf{Generated Model}}\\
		\cmidrule(lr){2-4} \cmidrule(l){5-7}
		& $\mathbf{R^2}$ & \textbf{\acs{MAE}} & \textbf{\acs{RMSE}} & $\mathbf{R^2}$ & \textbf{\acs{MAE}} & \textbf{\acs{RMSE}}\\
		\midrule
		
		%
		%
		\multirow{2}{0.5cm}{\textbf{\acs{ANN}}} & $0.807$ & $2.66$ & $3.731$ & $0.807$ & $2.66$ & $3.731$  \\
		& $\pm0.026$ & $\pm0.127$ & $\pm0.234$ & $\pm0.026$ & $\pm0.157$ & $\pm0.234$  \\
		
		%
		%
		\multirow{2}{0.5cm}{\textbf{\acs{M5}}} & $0.772$ & $2.773$ & $4.022$ & $0.771$ & $2.784$ & $4.033$ \\
		& $\pm0.03$ & $\pm0.81$ & $\pm0.206$ & $\pm0.03$ & $\pm0.77$ & $\pm0.206$ \\
		
		%
		%
		\multirow{2}{0.5cm}{\textbf{\acs{RF}}} & $0.829$ & $2.457$ & $3.485$ & $0.826$ & $2.475$ & $3.514$ \\ 
		& $\pm0.018$ & $\pm0.85$ & $\pm0.133$ & $\pm0.017$ & $\pm0.90$ & $\pm0.137$  \\
		
		%
		%
		\multirow{2}{0.5cm}{\textbf{\acs{SVM}}} & $0.552$ & $4.351$ & $5.666$ & $0.551$ & $4.36$ & $5.68$  \\
		& $\pm0.03$ & $\pm0.147$ & $\pm0.192$ & $\pm0.03$ & $\pm0.149$ & $\pm0.19$ \\

		\bottomrule
	\end{tabularx}
	\label{tab:validation_reg}
\end{table}

%% file: tex/tables/validation_class.tex
\newcolumntype{Y}{>{\centering\arraybackslash}X}

\begin{table}[ht]
	\centering
	\caption{Validation of the Code Generator for Classification Models}
	\vspace{-0.3cm}
	
	\begin{tabularx}{\columnwidth}{l c*{8}{Y}}
		\toprule
		\textbf{Model} & \multicolumn{4}{c}{\textbf{\acs{WEKA}}} & \multicolumn{4}{c}{\textbf{Generated Model}}\\
		\cmidrule(lr){2-5} \cmidrule(l){6-9}
		& \textbf{ACC} & \textbf{PREC} & \textbf{REC} & $\mathbf{F}_{1}$  & \textbf{ACC} & \textbf{PREC} & \textbf{REC} & $\mathbf{F}_{1}$ \\
		\midrule

		\multirow{2}{0.5cm}{\textbf{\acs{ANN}}} 
		& $93.79$ & $96.59$ & $97.19$ & $96.89$ & $93.79$ & $96.59$ & $97.19$ & $96.89$ \\
		& $\pm1.16$ & $\pm0.84$ & $\pm1.03$ & $\pm0.61$ & $\pm1.16$ & $\pm0.84$ & $\pm1.03$ & $\pm0.61$ \\

		\multirow{2}{0.5cm}{\textbf{\acs{RF}}} 
		& $93.67$ & $96.15$ & $97.31$ & $96.72$ & $93.52$ & $95.97$ & $97.29$ & $96.62$ \\
		& $\pm1.58$ & $\pm1.26$ & $\pm0.7$ & $\pm0.89$ & $\pm1.51$ & $\pm1.19$ & $\pm0.74$ & $\pm0.81$ \\

		\multirow{2}{0.5cm}{\textbf{\acs{SVM}}} 
		& $92.48$ & $94.97$ & $97.56$ & $96.24$ & $92.48$ & $95.05$ & $97.56$ & $96.29$ \\
		& $\pm0.62$ & $\pm0.79$ & $\pm0.55$ & $\pm0.43$ & $\pm0.66$ & $\pm0.85$ & $\pm0.56$ & $\pm0.49$ \\

		\bottomrule
	\end{tabularx}
	\label{tab:validation_class}
	
	\emph{ACC}: Accuracy, \emph{PREC}: Precision, \emph{REC}: Recall, \emph{$\text{F}_{1}$}: $\text{F}_{1}$ Score
\end{table}


%% file: tex/conclusion.tex
\section{Conclusion}

%
%
In this paper, we presented \ac{LIMITS}---a novel open source machine learning framework for \ac{IoT} applications, which provides automation features for high-level data analysis tasks and platform-specific code generation.
%
%
In contrast to existing solution approaches, \ac{LIMITS} explicitly integrates the platform-specific resource constraints and compilation toolchain of the target \ac{IoT} platform into the model selection process.
%
%
Its potential of catalyzing the development of machine learning-enabled \ac{IoT} systems was demonstrated based on two case studies focusing on cellular data rate prediction in vehicular networks and radio-based vehicle classification.
%
%
In future work, we will integrate further machine learning models into \ac{LIMITS}. Furthermore, we consider integrating automatic static \ac{WCET} analysis into the model selection process.

%% file: tex/acknowledgment.tex
\ifdoubleblind

\else
\section*{Acknowledgment}

\footnotesize
Part of the work on this paper has been supported by Deutsche Forschungsgemeinschaft (DFG) within the Collaborative Research Center SFB 876 ``Providing Information by Resource-Constrained Analysis'', projects B4. Parts of this research have been funded by the Federal Ministry of Education and Research of Germany as part of the competence center for machine learning ML2R (01$|$S18038A).

%% file: Manuscript.bbl
\begin{thebibliography}{10}
\providecommand{\url}[1]{#1}
\csname url@samestyle\endcsname
\providecommand{\newblock}{\relax}
\providecommand{\bibinfo}[2]{#2}
\providecommand{\BIBentrySTDinterwordspacing}{\spaceskip=0pt\relax}
\providecommand{\BIBentryALTinterwordstretchfactor}{4}
\providecommand{\BIBentryALTinterwordspacing}{\spaceskip=\fontdimen2\font plus
\BIBentryALTinterwordstretchfactor\fontdimen3\font minus
  \fontdimen4\font\relax}
\providecommand{\BIBforeignlanguage}[2]{{%
\expandafter\ifx\csname l@#1\endcsname\relax
\typeout{** WARNING: IEEEtran.bst: No hyphenation pattern has been}%
\typeout{** loaded for the language `#1'. Using the pattern for}%
\typeout{** the default language instead.}%
\else
\language=\csname l@#1\endcsname
\fi
#2}}
\providecommand{\BIBdecl}{\relax}
\BIBdecl

\bibitem{Zanella/etal/2014a}
A.~{Zanella}, N.~{Bui}, A.~{Castellani}, L.~{Vangelista}, and M.~{Zorzi},
  ``Internet of things for smart cities,'' \emph{IEEE Internet of Things
  Journal}, vol.~1, no.~1, pp. 22--32, Feb 2014.

\bibitem{Terroso-Saenz/etal/2019a}
F.~Terroso-Saenz, A.~González-Vidal, A.~P. Ramallo-González, and A.~F.
  Skarmeta, ``An open {IoT} platform for the management and analysis of energy
  data,'' \emph{Future Generation Computer Systems}, vol.~92, pp. 1066 -- 1079,
  2019.

\bibitem{Sliwa/etal/2019b}
B.~Sliwa, T.~Liebig, T.~Vranken, M.~Schreckenberg, and C.~Wietfeld,
  ``System-of-systems modeling, analysis and optimization of hybrid vehicular
  traffic,'' in \emph{2019 Annual IEEE International Systems Conference
  (SysCon)}, Orlando, Florida, USA, Apr 2019.

\bibitem{Yang/etal/2019a}
P.~{Yang}, Y.~{Xiao}, M.~{Xiao}, and S.~{Li}, ``{6G} wireless communications:
  {Vi}sion and potential techniques,'' \emph{IEEE Network}, vol.~33, no.~4, pp.
  70--75, July 2019.

\bibitem{Bui/etal/2017a}
N.~Bui, M.~Cesana, S.~A. Hosseini, Q.~Liao, I.~Malanchini, and J.~Widmer, ``A
  survey of anticipatory mobile networking: Context-based classification,
  prediction methodologies, and optimization techniques,'' \emph{IEEE
  Communications Surveys \& Tutorials}, 2017.

\bibitem{Liang/etal/2019a}
L.~{Liang}, H.~{Ye}, and G.~Y. {Li}, ``Toward intelligent vehicular networks:
  {A} machine learning framework,'' \emph{IEEE Internet of Things Journal},
  vol.~6, no.~1, pp. 124--135, Feb 2019.

\bibitem{Ye/etal/2018a}
H.~Ye, L.~Liang, G.~Y. Li, J.~Kim, L.~Lu, and M.~Wu, ``Machine learning for
  vehicular networks: {R}ecent advances and application examples,'' \emph{IEEE
  Vehicular Technology Magazine}, vol.~13, no.~2, pp. 94--101, June 2018.

\bibitem{Breiman/2001a}
L.~Breiman, ``Random forests,'' \emph{Machine Learning}, vol.~45, no.~1, pp.
  5--32, 2001.

\bibitem{Quinlan/1992a}
J.~R. Quinlan, ``Learning with continuous classes.''\hskip 1em plus 0.5em minus
  0.4em\relax World Scientific, 1992, pp. 343--348.

\bibitem{Cortes/Vapnik/1995a}
C.~Cortes and V.~Vapnik, ``Support-vector networks,'' \emph{Machine Learning},
  vol.~20, no.~3, pp. 273--297, Sep. 1995.

\bibitem{Goodfellow/etal/2016a}
I.~Goodfellow, Y.~Bengio, and A.~Courville, \emph{Deep Learning}.\hskip 1em
  plus 0.5em minus 0.4em\relax MIT Press, 2016,
  \url{http://www.deeplearningbook.org}.

\bibitem{Sliwa/etal/2018b}
B.~Sliwa, T.~Liebig, R.~Falkenberg, J.~Pillmann, and C.~Wietfeld, ``Efficient
  machine-type communication using multi-metric context-awareness for cars used
  as mobile sensors in upcoming {5G} networks,'' in \emph{2018 IEEE 87th
  Vehicular Technology Conference (VTC-Spring)}, Porto, Portugal, Jun 2018,
  {Best Student Paper Award}.

\bibitem{Sliwa/Wietfeld/2019b}
B.~Sliwa and C.~Wietfeld, ``Empirical analysis of client-based network quality
  prediction in vehicular multi-{MNO} networks,'' in \emph{2019 IEEE 90th
  Vehicular Technology Conference (VTC-Fall)}, Honolulu, Hawaii, USA, Sep 2019.

\bibitem{Sliwa/Wietfeld/2020a}
------, ``A reinforcement learning approach for efficient opportunistic
  vehicle-to-cloud data transfer,'' in \emph{2020 IEEE Wireless Communications
  and Networking Conference (WCNC)}, Seoul, South Korea, Apr 2020.

\bibitem{Sliwa/etal/2019a}
B.~Sliwa, R.~Falkenberg, T.~Liebig, N.~Piatkowski, and C.~Wietfeld, ``Boosting
  vehicle-to-cloud communication by machine learning-enabled context
  prediction,'' \emph{IEEE Transactions on Intelligent Transportation Systems},
  Jul 2019.

\bibitem{Falkenberg/etal/2018a}
R.~Falkenberg, B.~Sliwa, N.~Piatkowski, and C.~Wietfeld, ``Machine learning
  based uplink transmission power prediction for {LTE} and upcoming {5G}
  networks using passive downlink indicators,'' in \emph{2018 IEEE 88th
  Vehicular Technology Conference (VTC-Fall)}, Chicago, USA, Aug 2018.

\bibitem{Sliwa/Wietfeld/2019c}
B.~Sliwa and C.~Wietfeld, ``Towards data-driven simulation of end-to-end
  network performance indicators,'' in \emph{2019 IEEE 90th Vehicular
  Technology Conference (VTC-Fall)}, Honolulu, Hawaii, USA, Sep 2019.

\bibitem{Sliwa/etal/2020b}
B.~Sliwa, R.~Falkenberg, and C.~Wietfeld, ``Towards cooperative data rate
  prediction for future mobile and vehicular 6g networks,'' in \emph{2nd 6G
  Wireless Summit (6G SUMMIT)}.\hskip 1em plus 0.5em minus 0.4em\relax Levi,
  Finland: IEEE, Mar 2020.

\bibitem{Park/etal/2016a}
T.~{Park}, N.~{Abuzainab}, and W.~{Saad}, ``Learning how to communicate in the
  internet of things: Finite resources and heterogeneity,'' \emph{IEEE Access},
  vol.~4, pp. 7063--7073, 2016.

\bibitem{Yao/etal/2018a}
S.~Yao, Y.~Zhao, H.~Shao, S.~Liu, D.~Liu, L.~Su, and T.~Abdelzaher,
  ``{FastDeepIoT}: {T}owards understanding and optimizing neural network
  execution time on mobile and embedded devices,'' in \emph{Proceedings of the
  16th ACM Conference on Embedded Networked Sensor Systems}, ser. SenSys '18,
  New York, NY, USA, 2018, pp. 278--291.

\bibitem{Masoudinejad/etal/2018a}
M.~{Masoudinejad}, A.~K. {Ramachandran Venkatapathy}, D.~{Tondorf},
  D.~{Heinrich}, R.~{Falkenberg}, and M.~{Buschhoff}, ``Machine learning based
  indoor localisation using environmental data in {PhyNetLab} warehouse,'' in
  \emph{Smart SysTech 2018; European Conference on Smart Objects, Systems and
  Technologies}, June 2018, pp. 1--8.

\bibitem{Bucila/etal/2006a}
C.~Bucila, R.~Caruana, and A.~Niculescu{-}Mizil, ``Model compression,'' in
  \emph{Proceedings of the Twelfth {ACM} {SIGKDD} International Conference on
  Knowledge Discovery and Data Mining, Philadelphia, PA, USA, August 20-23,
  2006}, 2006, pp. 535--541.

\bibitem{Piatkowski/etal/2013a}
N.~Piatkowski, S.~Lee, and K.~Morik, ``Spatio-temporal random fields:
  compressible representation and distributed estimation,'' \emph{Machine
  Learning}, vol.~93, no.~1, pp. 115--139, 2013.

\bibitem{Kumar/etal/2017a}
A.~Kumar, S.~Goyal, and M.~Varma, ``Resource-efficient machine learning in 2
  {KB} {RAM} for the internet of things,'' in \emph{Proceedings of the 34th
  International Conference on Machine Learning}, ser. Proceedings of Machine
  Learning Research, D.~Precup and Y.~W. Teh, Eds., vol.~70.\hskip 1em plus
  0.5em minus 0.4em\relax International Convention Centre, Sydney, Australia:
  PMLR, 06--11 Aug 2017, pp. 1935--1944.

\bibitem{Shukla/Munir/2017a}
R.~M. {Shukla} and A.~{Munir}, ``An efficient computation offloading
  architecture for the internet of things (iot) devices,'' in \emph{2017 14th
  IEEE Annual Consumer Communications Networking Conference (CCNC)}, Jan 2017,
  pp. 728--731.

\bibitem{Piatkowski/etal/2016a}
N.~Piatkowski, S.~Lee, and K.~Morik, ``Integer undirected graphical models for
  resource-constrained systems,'' \emph{Neurocomputing}, vol. 173, pp. 9 -- 23,
  2016.

\bibitem{Hofmann/Klinkenberg/2013a}
M.~Hofmann and R.~Klinkenberg, \emph{{RapidMiner}: {D}ata mining use cases and
  business analytics applications}.\hskip 1em plus 0.5em minus 0.4em\relax
  Chapman \& Hall/CRC, 2013.

\bibitem{Berthold/etal/2009a}
M.~R. Berthold, N.~Cebron, F.~Dill, T.~R. Gabriel, T.~K\"{o}tter, T.~Meinl,
  P.~Ohl, K.~Thiel, and B.~Wiswedel, ``{KNIME} - the konstanz information
  miner: {V}ersion 2.0 and beyond,'' \emph{SIGKDD Explor. Newsl.}, vol.~11,
  no.~1, pp. 26--31, Nov. 2009.

\bibitem{Hall/etal/2009a}
M.~Hall, E.~Frank, G.~Holmes, B.~Pfahringer, P.~Reutemann, and I.~H. Witten,
  ``The {WEKA} data mining software: {A}n update,'' \emph{SIGKDD Explorations},
  vol.~11, no.~1, pp. 10--18, 2009.

\bibitem{Pedregosa/etal/2011a}
F.~Pedregosa, G.~Varoquaux, A.~Gramfort, V.~Michel, B.~Thirion, O.~Grisel,
  M.~Blondel, P.~Prettenhofer, R.~Weiss, V.~Dubourg, J.~Vanderplas, A.~Passos,
  D.~Cournapeau, M.~Brucher, M.~Perrot, and E.~Duchesnay, ``Scikit-learn:
  {M}achine learning in python,'' \emph{Journal of Machine Learning Research},
  vol.~12, pp. 2825--2830, 2011.

\bibitem{Paszke/etal/2017a}
A.~Paszke, S.~Gross, S.~Chintala, G.~Chanan, E.~Yang, Z.~DeVito, Z.~Lin,
  A.~Desmaison, L.~Antiga, and A.~Lerer, ``Automatic differentiation in
  {PyTorch},'' 2017.

\bibitem{Abadi/etal/2016a}
M.~Abadi, P.~Barham, J.~Chen, Z.~Chen, A.~Davis, J.~Dean, M.~Devin,
  S.~Ghemawat, G.~Irving, M.~Isard, M.~Kudlur, J.~Levenberg, R.~Monga,
  S.~Moore, D.~G. Murray, B.~Steiner, P.~Tucker, V.~Vasudevan, P.~Warden,
  M.~Wicke, Y.~Yu, and X.~Zheng, ``{TensorFlow}: {A} system for large-scale
  machine learning,'' in \emph{12th USENIX Symposium on Operating Systems
  Design and Implementation (OSDI 16)}, 2016, pp. 265--283.

\bibitem{Jomrich/etal/2018a}
F.~Jomrich, A.~Herzberger, T.~Meuser, B.~Richerzhagen, R.~Steinmetz, and
  C.~Wille, ``Cellular bandwidth prediction for highly automated driving -
  {E}valuation of machine learning approaches based on real-world data,'' in
  \emph{Proceedings of the 4th International Conference on Vehicle Technology
  and Intelligent Transport Systems 2018}, no.~4.\hskip 1em plus 0.5em minus
  0.4em\relax SCITEPRESS, Mar 2018, pp. 121--131.

\bibitem{Samba/etal/2017a}
A.~Samba, Y.~Busnel, A.~Blanc, P.~Dooze, and G.~Simon, ``Instantaneous
  throughput prediction in cellular networks: {W}hich information is needed?''
  in \emph{2017 IFIP/IEEE Symposium on Integrated Network and Service
  Management (IM)}, May 2017, pp. 624--627.

\bibitem{Apajalahti/etal/2018a}
K.~{Apajalahti}, E.~A. {Walelgne}, J.~{Manner}, and E.~{Hyvönen},
  ``Correlation-based feature mapping of crowdsourced {LTE} data,'' in
  \emph{2018 IEEE 29th Annual International Symposium on Personal, Indoor and
  Mobile Radio Communications (PIMRC)}, Sep. 2018, pp. 1--7.

\bibitem{Louppe/etal/2013a}
G.~Louppe, L.~Wehenkel, A.~Sutera, and P.~Geurts, ``Understanding variable
  importances in forests of randomized trees,'' in \emph{Proceedings of the
  26th International Conference on Neural Information Processing Systems -
  Volume 1}, ser. NIPS'13.\hskip 1em plus 0.5em minus 0.4em\relax USA: Curran
  Associates Inc., 2013, pp. 431--439.

\bibitem{Sliwa/etal/2018e}
B.~Sliwa, N.~Piatkowski, M.~Haferkamp, D.~Dorn, and C.~Wietfeld, ``Leveraging
  the channel as a sensor: Real-time vehicle classification using
  multidimensional radio-fingerprinting,'' in \emph{2018 IEEE 21st
  International Conference on Intelligent Transportation Systems (ITSC)}, Maui,
  Hawaii, USA, Nov 2018.

\end{thebibliography}
